\RequirePackage{fix-cm}
\documentclass[twocolumn,epjc3]{svjour3}  
\smartqed  % flush right qed marks, e.g. at end of proof
\RequirePackage{graphicx}
\journalname{Eur. Phys. J. C}
\usepackage{cite} %A: para contraer referencias
\usepackage{amsmath}%A
\usepackage{hyperref}
\usepackage{cite}%A
\usepackage{amsmath,amssymb}%A
\usepackage{xcolor}%A

\begin{document}

\title{Scale-dependent (2+1) - dimensional electrically charged black holes in Einstein-power-Maxwell theory}

\author{\'Angel Rinc\'on \thanksref{e1,addr1}
        \and
        Ernesto Contreras \thanksref{
        e2,addr2,ol} %etc.
        \and
        Pedro Bargue\~no \thanksref{e3,addr3}
        \and
        Benjamin Koch \thanksref{e4,addr1}
        \and
        Grigorios Panotopoulos \thanksref{e5,addr4}          
     }

\thankstext{e1}{e-mail: \href{mailto:arrincon@uc.cl}{\nolinkurl{arrincon@uc.cl}} }
\thankstext{e2}{e-mail: \href{mailto:ejcontre@espol.edu.ec}{\nolinkurl{ejcontre@espol.edu.ec}} }
\thankstext{e3}{e-mail: \href{mailto:p.bargueno@uniandes.edu.co}{\nolinkurl{p.bargueno@uniandes.edu.co }} }
\thankstext{e4}{e-mail: \href{mailto:bkoch@fis.puc.cl}{\nolinkurl{bkoch@fis.puc.cl}} }
\thankstext{e5}{e-mail: \href{mailto:grigorios.panotopoulos@tecnico.ulisboa.pt}{\nolinkurl{grigorios.panotopoulos@tecnico.ulisboa.pt}} }

\thankstext{ol}{On leave from Universidad Central de Venezuela}
\authorrunning{Rinc\'on et al.} % if too long for running head

\institute{Instituto de F{\'i}sica, Pontificia Universidad Cat{\'o}lica de Chile,\\ Av. Vicu{\~n}a Mackenna 4860, Santiago, Chile. \label{addr1}
           \and
Facultad de Ciencias Naturales y Matem\'aticas, Escuela Superior Polit\'ecnica del Litoral, ESPOL, 
\\
Apartado Postal 09-01-5863, 
Campus Gustavo Galindo Km 30.5 V\'ia Perimetral, Guayaquil, Ecuador.
\label{addr2}
           \and
Departamento de F\'isica, Universidad de los Andes, \\
Apartado A\'ereo 4976, Bogot\'a, Distrito Capital, Colombia \label{addr3}
           \and
           Centro de Astrof{\'i}sica e Gravita{\c c}{\~a}o, Instituto Superior T{\'e}cnico-IST, \\
           Universidade de Lisboa-UL, Av. Rovisco Pais, 1049-001 Lisboa, Portugal\label{addr4}        
}

\date{Received: date / Accepted: date}

\maketitle

\begin{abstract}
In this work we extend and generalize our previous work on the scale dependence at the level of the effective action of  black holes in the presence of non-linear electrodynamics. In particular, we consider the Einstein-power-Maxwell theory without a cosmological constant in (2+1) dimensions, assuming a scale dependence of both the gravitational and the electromagnetic coupling and we investigate in detail how the scale--dependent scenario affects the horizon and thermodynamic properties of the classical black holes for any value of the power parameter. 
In addition, we solve the corresponding effective field equations imposing the ``null energy condition" in order to obtain analytical solutions. 
The implications of quantum corrections are also briefly discussed.

%Include keywords, PACS and mathematical subject classification numbers as needed.
%\PACS{PACS code1 \and PACS code2 \and more}
\end{abstract}

%%%%%%%%%%%%%%%%%%%%%
\section{Introduction}

Three-dimensional gravity is attracting a lot of attention for several reasons. On one hand due to the deep connection to Yang-Mills and Chern-Simons theory \cite{CS,witten1,witten2}. On the other hand in this lower dimensional gravitational theory, there are no propagating degrees of freedom, which makes analytic manipulations much more accessible. 
Furthermore, three-dimensional black holes are characterized by properties also found in their four-dimensional counterparts, such as horizon radius, temperature, entropy etc. Therefore, three-dimensional gravity allows to get deep insight into the corresponding systems   that live in four-dimensions.   

The main motivation to study non-linear electrodynamics (NLED) was to overcome certain problems present in 
the standard Maxwell's theory. Initially, the so called Born-Infeld non-linear electrodynamics was introduced in the 30's in order 
to obtain a finite self-energy of point-like charges \cite{BI}. During the last decades, these type of 
models reappear
in the open sector of superstring theories \cite{stringtheory} as their describe the dynamics of D-branes \cite{Dbranes}. 
Similarly, in heterotic string theory a Gauss-Bonnet term coupled to quartic contractions of the Maxwell field strength appears. \cite{Zwiebach:1985uq,Corley:2001hg,Kats:2006xp,Anninos:2008sj,Cai:2008ph}.

Also, this kind of electrodynamics has been coupled to gravity in order to obtain, for example, regular black hole  solutions \cite{ayon1998,ayon1998a, bargueno2016b}, semiclassical corrections to the black hole entropy \cite{bargueno2016a}, 
and novel exact solutions with a
cosmological constant acting as an effective Born-Infeld cut-off \cite{BarguenoVagenas2016}.

A particularly interesting class of 
NLED theories is the so called power-Maxwell theory (EpM hereafter).
There are several reasons to study 
the Einstein-power-Maxwell electrodynamics, as it was recently pointed out in \cite{Gurtug:2010dr}: "In recent years, the use of power Maxwell fields has attracted considerable interest. It has been used for obtaining solutions in d-spacetime dimensions \cite{Hassaine:2008pw}, Ricci flat rotating black branes with a conformally Maxwell source \cite{Hendi:2009zzb}, Lovelock black holes \cite{Maeda:2008ha}, Gauss-Bonnet gravity \cite{Hendi:2010zza}, and the effect of power Maxwell field on the magnetic solutions in Gauss-Bonnet gravity \cite{Hendi:2010dz}."

The EpM theory is
described by a Lagrangian density of the form $\mathcal{L}(F)=F^{\beta}$, where $F=F_{\mu \nu}F^{\mu \nu}/4$ is the Maxwell 
invariant, and $\beta$ is an arbitrary rational number. When $\beta=1$ one  recovers the standard linear electrodynamics, 
while for $\beta=D/4$, with $D$ being the dimensionality of space time, the electromagnetic energy momentum tensor is 
traceless \cite{cataldo2000a, chinos}. In three dimensions 
the generic black hole solution without imposing the traceless condition has been found in \cite{Gurtug:2010dr}, while black hole 
solutions in linear Einstein-Maxwell theory are given in \cite{solution2,solution3}. 
Other interesting solutions and properties of black holes in the presence of 
power-Maxwell theory have been found in~\cite{Hassaine:2007py,Hassaine:2008pw,Gonzalez:2009nn,
Hendi:2010zza,Panotopoulos:2017hns,Panotopoulos:2018pvu}, 
whereas some topological black hole solutions with power-law Maxwell fields have been 
investigated in~\cite{Zangeneh:2015wia,Zangeneh:2015uwa,Zangeneh:2015gja}, as well as Born-Infeld theory in \cite{Panotopoulos:2017yoe,Destounis:2018utr}.
Interesting features arise from a study of the thermodynamic properties of EpM black holes, as discussed in \cite{Gonzalez:2009nn}.

It is well-known that one of the open issues in modern theoretical physics is a consistent formulation of quantum gravity. Although there are several approaches to the problem (for an incomplete list see 
e.g. \cite{QG1,QG2,QG3,QG4,QG5,QG6,QG7,QG8,QG9} and references therein), 
most of them have something in common, namely that the basic parameters 
that enter into the action, such as Newton's constant, the cosmological constant or the electromagnetic coupling, 
become scale--dependent quantities. As scale dependence at the level of the effective action is a generic result of 
quantum field theory, the resulting effective action of scale--dependent gravity is expected to modify the 
properties of classical black hole backgrounds.

It is the aim of this work to study the scale dependence at the level of the effective action of three-dimensional
charged black holes in the presence of the Einstein-power-Maxwell non-linear electrodynamics for any value of the power parameter, 
extending and generalizing  previous work \cite{previous}, where we imposed the traceless condition $\beta=3/4$. 
We will use the formalism and notation of \cite{previous}.

Our work is organized as follows. 
After this introduction we present the model and the field equations. Section \ref{Clasico}
is devoted to introduce the classical black hole background. 
In sections \ref{scale_setting} and \ref{NEC} we allow for scale dependent couplings, 
we impose the ``null energy condition", and after that we present our 
solution for the metric lapse function as well as for the couplings in the scale dependent scenario. 
In section \ref{Conclusions} we briefly discuss our main findings, concluding in the same section. 

%%%%%%%%%%%%%%%%%%%%%%%%%%%%%%
\section{Classical Einstein-power-Maxwell theory }

In this section we will present the classical theory of non--linear electrodynamics in (2+1) dimensional spacetimes for an arbitrary EpM theory (namely, for an arbitrary index $\beta$). Those theories will then be investigated in the context of scale--dependent couplings. 
Our starting point is the so-called Einstein-power-Maxwell action without cosmological constant $(\Lambda_0 =0)$, 
assuming the EpM Lagrangian density, i.e. $\mathcal{L}(F) = \gamma |F|^{\beta}$, which reads
\begin{align}\label{act1}
I_0[g_{\mu \nu}, A_{\mu}] &=  \int {\mathrm {d}}^{3}x {\sqrt {-g}}\,
\bigg[\frac{1}{2\kappa_0} R - \frac{1}{e_{0}^{2\beta}}\mathcal{L}(F) \bigg],
\end{align}
where $\kappa_0 \equiv 8\pi G_0$ is the gravitational coupling, 
$G_0$ is Newton's constant, 
$e_0$ is the electromagnetic coupling constant, 
$R$ is the Ricci scalar, $\mathcal{L}(F)$ is the electromagnetic Lagrangian density, $\gamma$ is a proportionality constant, 
$F$ is the Maxwell invariant previously defined, and
$F_{\mu \nu} = \partial_{\mu}A_{\nu} - \partial_{\nu}A_{\mu}$ 
is the electromagnetic field strength tensor.
We use the metric signature $(-, +, +)$, and natural units ($c = \hbar = k_B = 1$)
such that the action is dimensionless. Note that $\beta$ 
is an arbitrary rational number, which also appears in the exponent of the electromagnetic coupling in order to maintain the action dimensionless.
It is easy to check that the special case $\beta = 1$ reproduces the 
classical Einstein-Maxwell 
action, and thus 
the standard electrodynamics is recovered. For $\beta \neq 1$ one can obtain Maxwell-like solutions.
In the following we shall consider the general case, so that $\beta$ is taken to be a free parameter.
As our solution should reproduce the classical one, we restrict the values of this parameter by demanding the energy conditions to be satisfied. According to \cite{Gurtug:2010dr}, we will only take into account the (naive) range $\beta \in \Re^+$ (our solution, however, could have additional forbidden values of the parameter $\beta$).
The classical equations of motion for the metric field are given 
by Einstein's field equations
\begin{align}\label{Gmunu}
G_{\mu \nu} &= \frac{\kappa_0}{e_0^{2\beta}} T_{\mu \nu }.
\end{align}
The energy momentum tensor $T_{\mu \nu}$ is associated to the electromagnetic field strength $F_{\mu \nu}$ through
\begin{align}\label{TNL}
T_{\mu \nu} &\equiv T_{\mu \nu}^{\text{EM}} = \ 
\mathcal{L}(F) g_{\mu \nu} - \mathcal{L}_{F} F_{\mu \gamma} F_{\nu}\ ^{\gamma},
\end{align}
remembering that $\mathcal{L}_F= \mathrm{d}\mathcal{L}/\mathrm{d}F$. Besides, for static 
circularly symmetric solutions the electric 
field $E(r)$ is given by 
\begin{align}\label{Fmunu}
F_{\mu \nu} &= (\delta_{\mu}^{r} \delta_{\nu}^{t} - \delta_{\nu}^{r} \delta_{\mu}^{t})E(r).
\end{align}
Taking the variation of the classical action with respect to 
the field $A_{\mu}(x)$ one obtains
\begin{align} \label{decov}
D_{\mu}\left(\frac{\mathcal{L}_{F}F^{\mu\nu}}{e_0^{2\beta}}\right)& = 0,
\end{align}
where $e_0^{2\beta}$ is a constant.
Combining Eq. \eqref{Gmunu} with Eq. \eqref{decov} we are able to 
determine the set of functions $\{f(r), E(r)\}$.  
It should be noted that the general solution of this problem was previously appointed in Ref. \cite{Gurtug:2010dr} by
computing the lapse function and the electric field, as well as the corresponding thermodynamic properties. 

%%%%%%%%%%%%%%%%%%%%%%%%%%%%%%%%
\section{Black hole solution for Einstein-Maxwell model of arbitrary power}
\label{Clasico}
The general metric ansatz assuming circular symmetry is given by
%For the metric spherical symmetry implies
\begin{equation}\label{metric}
\mathrm{d}s^2 = -f(r) \mathrm {d}t^2 + g(r) \mathrm {d}r^2 + r^2 \mathrm {d} \phi^2.
\end{equation}
Note that, in the classical solution, it is possible to deduce the Schwarzschild relation, namely $g(r)=f(r)^{-1}$. 
The classical (2+1)-dimensional Einstein-Maxwell black
hole solution (for an arbitrary index $\beta$) is obtained after 
solving $f_0(r)$ and $E_0(r)$ and was previously found in Ref. \cite{Gurtug:2010dr}. As we will compare these results with the scale-dependent solution provided in Section \ref{solution},
 here we will briefly comment the main features of the classical case.
Then, solving the Einstein field equations we obtain:
\begin{align}
f_0(r) & = B r^{1-\alpha }+\frac{C}{\alpha -1},
\\
E_0(r) &= A \left[ \frac{e_0^{\alpha +1}}{r^{\alpha }} \right].
\end{align}
where the set $\{A, B, C\}$ are constants of integration which must be fixed. According to 
Ref. \cite{Gurtug:2010dr}, the parameter $C$ is related with the mass of the 
black hole $M_0$ while $B$ takes into account the classical charge $Q_0$ (the same for the parameter $A$). 
In addition, note that the auxiliary parameter $\alpha$ is defined as follow:
\begin{align}
\alpha = \ &\frac{1}{2\beta-1}.
\end{align}
The next step consists in computing the horizon of this black hole, which is
\begin{align}
r_0 &= \left(\frac{C}{B(1-\alpha)}\right)^{\frac{1}{1-\alpha }}.
\end{align}
By writing the lapse function in terms of the classical horizon we have
\begin{align}
f_0(r) &= \frac{C}{\alpha -1} \left[1-\left(\frac{r_0}{r}\right)^{\alpha -1}\right].
\end{align}
Another important point is the thermodynamics of the system. We can then define three quantities, i. e.,  
the Hawking temperature, $T_H$, the Bekenstein-Hawking entropy, $S$, and the specific heat, $C_Q$. Their 
corresponding expressions are given by
\begin{align}
T_0(r_0) &= \frac{1}{4\pi} \left|\ \frac{C}{r_0} \ \right|,
\\
S_0(r_0) &= \frac{\mathcal{A}_0}{4 G_0 },
\\
C_0(r_0) &= T \ \frac{\partial S}{\partial T} \ \Bigg{|}_{r_0},
\end{align}
being $\mathcal{A}_0$ the horizon area defined as
\begin{align}
\mathcal{A}_{0} &= \oint \mathrm{d}x \sqrt{h} = 2\pi r_0.
\end{align}
where $h_{ij}$ is the induced metric at the horizon $r_0$.
%%%%%%%%%%%%%%%%%%%%%%%%%%%%%%%%%
\section{Scale dependent coupling and scale setting}
\label{scale_setting}
This section summarizes the equations of motion for the scale--dependent Einstein-power-Maxwell theory with arbitrary index. 
The idea and notation follows \cite{previous,Koch:2014joa,angel,Rincon:2017ypd,Rincon:2017ayr,Hernandez-Arboleda:2018qdo,Contreras:2017eza,Contreras:2018dhs,Contreras:2018swc,Rincon:2018lyd}.
\noindent The scale--dependent couplings of the theory are i) the Newton's coupling $G_k$ 
(which can be related with the gravitational coupling by $\kappa_k \equiv 8 \pi G_k$), and ii) 
the electromagnetic coupling $1/e_k$. Furthermore, there are three independent fields, which are the metric $g_{\mu \nu}(x)$, the electromagnetic
four-potential $A_{\mu}(x)$, and the  scale field $k(x)$. 
The effective action for this theory reads
\begin{align}\label{Effective_Actiom}
\Gamma[g_{\mu \nu}, A_{\mu}, k] &=  \int {\mathrm {d}}^{3}x {\sqrt {-g}}\,
\bigg[\frac{1}{2\kappa_k} R - \frac{1}{e_{k}^{2\beta}}\mathcal{L}(F) \bigg].
\end{align}
The equations of motion obtained from a variation of (\ref{Effective_Actiom}) with respect to
$g_{\mu \nu}(x)$ are
\begin{align}
G_{\mu\nu} &= \frac{\kappa_k}{e^{2\beta}_k}T^{\text{eff}}_{\mu\nu},
\end{align}
where
\begin{align}
T^{\text{eff}}_{\mu\nu} &= T^{\text{EM}}_{\mu\nu} - \frac{e^{2\beta}_k}{\kappa_k} \Delta t_{\mu \nu}.
\end{align}
Note that $T^{\text{EM}}_{\mu\nu}$ is given by (\ref{TNL})
and the additional contribution $\Delta t_{\mu \nu}$ is 
\begin{align}
\Delta t_{\mu\nu} &= G_k \Bigl(g_{\mu \nu} \square - \nabla_{\mu} \nabla_{\nu}
\Bigl)G_k^{-1}.
\end{align}
The equations of motion for
the four-potential $A_{\mu}(x)$ taking into account the running of $e_k$ are
\begin{align} \label{decovcoupling}
D_{\mu}\left(\frac{\mathcal{L}_{F}F^{\mu\nu}}{e_k^{2\beta}}\right)& = 0.
\end{align}
It is important to note that, 
in any quantum field theory the renormalization scale $k$
has to be set to a quantity characterizing the physical system under consideration.
Thus, for background solutions of the gap equations, 
it is not constant anymore.
However, having an arbitrarily chosen non-constant $k=k(x)$ implies that
the set of equations of motion does not close consistently. 
This implies that the stress energy tensor is most likely not conserved
for almost any choice of the functional dependence $k=k(x)$.
This type of scenario has been largely explored in the context
of renormalization group improvement of black holes in asymptotic safety scenarios \cite{Bonanno:2000ep,Bonanno:2006eu,Reuter:2006rg,Reuter:2010xb,Falls:2012nd,Cai:2010zh,Becker:2012js,Becker:2012jx,Koch:2013owa,Koch:2013rwa,Ward:2006vw,Burschil:2009va,Falls:2010he,Koch:2014cqa,Bonanno:2016dyv}.
The loss of a conservation laws comes from the fact that there is one consistency equation missing.
This missing equation can be obtained from varying the effective action (\ref{Effective_Actiom}) with respect to the scale field $k(r)$, i.e.
\begin{equation}\label{vary}
\frac{\mathrm{d}}{\mathrm{d} k} \Gamma[g_{\mu \nu}, A_{\mu}, k]=0,
\end{equation}
which can thus be understood
as variational scale setting procedure \cite{Reuter:2003ca,Koch:2010nn,Domazet:2012tw,Koch:2014joa,Contreras:2016mdt}.
The combination of (\ref{vary}) with the above equations of motion guarantees the conservation of the stress energy tensor.
A detailed analysis of the split symmetry within the functional renormalization group equations supports this approach of dynamic scale setting \cite{Percacci:2016arh}.
To apply the variational procedure (\ref{vary}), however, the knowledge of the exact beta functions of the problem is required.
Since in many cases the precise form of these functions 
is unknown (or at least uncertain) one can, for
the case of simple black holes, impose a null energy condition
and solve for the couplings $G(r),\, \Lambda(r),\, e(r)$ directly 
\cite{Contreras:2013hua,Koch:2015nva,angel,Rincon:2017ypd,Contreras:2017eza,
Contreras:2018dhs,Rincon:2018sgd}.
This philosophy of assuring the consistency of the equations by imposing
a null energy condition will also be applied in the following study on
Einstein-Maxwell and Einstein-power-Maxwell black holes.

%%%%%%%%%%%%%%%%%%%%%%%%%%
\section{The null energy condition}\label{NEC}

An energy condition is, basically, an additional relation one imposes on the matter stress-energy tensor e.g. in order to try to capture 
the idea that ``energy should be positive"~\cite{Curiel:2014zba}. 
There are typically four energy conditions 
(dominant, weak, strong, and null) which help 
to obtain desirable solutions of Einstein's field 
equations \cite{Rubakov:2014jja,Wald:1984rg}.
Among those conditions, the null energy condition (NEC) is particularly interesting
since it is a crucial assumption of the Penrose(-Hawking) singularity theorem \cite{Penrose:1964wq}, valid in General Relativity.
Thus, for matter obeying the NEC, there is always a singularity that gets formed inside a black hole horizon, and any contracting Universe ends up in a singularity, provided its spatial curvature is dynamically negligible \cite{Rubakov:2014jja}.
Thus, we will focus our attention to the NEC. Our starting point is to consider certain null vector, called $\ell^{\mu}$, and to contract it with the matter stress energy tensor as 
NEC demands, i.e. : 
\begin{align}
T^{m}_{\mu \nu} \ell^{\mu} \ell^{\nu} \geq 0.
\end{align}
This ``trick" was used in Ref. \cite{angel} inspired by the Jacobson idea \cite{Jacobson:2007tj} on getting 
acceptable physical solutions. Note that in proving fundamental black hole theorems, such as the no hair theorem \cite{Heusler:1996ft}, and the 
second law of black hole thermodynamics \cite{Bardeen:1973gs}, the NEC is, indeed, required.
In the scale dependent scenario, we maintain the same condition in a more restrictive and thus more useful form
by making the inequality an equality 
\begin{align}\label{NEC1}
T^{\text{eff}}_{\mu \nu} \ell^{\mu} \ell^{\nu} = \bigg(T^{\text{EM}}_{\mu \nu} - \frac{e^{2\beta}_k}{\kappa_k} \Delta t_{\mu \nu}\bigg) \ell^{\mu} \ell^{\nu} = 0.
\end{align}
For the null vector we choose a
radial null vector $\ell^{\mu}=\{f^{-1/2}, f^{1/2}, 0\}$.
Since the electromagnetic contribution to the effective stress energy tensor (\ref{TNL})
satisfies the NEC (\ref{NEC1}) by construction, the same has to hold
for the additional contribution introduced due to the scale dependence of the gravitiational coupling i.e. 
\begin{align}\label{Delta_t}
\Delta t_{\mu \nu} \ell^{\mu} \ell^{\nu} = 0.
\end{align}
%
%

%%%%%%%%%%%%%%%%%%%%%%%%%%%%%%%
\section{Scale dependent Einstein-power-Maxwell theory}

%%%%%%%%%%%%%%%%%
\label{solution}
\subsection{Solution}
In order to obtain the full solution with circular symmetry, we need to find the set $\{G(r), E(r), f(r), e(r)^{\alpha+1}\}$. 
We first start by considering the constraint given by the NEC. The Eq. \eqref{Delta_t} gives an explicit differential equation for the gravitational coupling $G(r)$, i.e.
\begin{align}
G(r)\frac{{\mathrm {d}}^2 G(r)}{\mathrm {d} r^2} - 2 \left(\frac{\mathrm {d} G(r)}{\mathrm {d} r}\right)^2=0,
\end{align}
which allows us to obtain 
\begin{align}\label{Gr}
G(r) = \frac{G_0}{1 + \epsilon r}.
\end{align}
After that, we use equation of motion for the 4-potential given by Eq. \eqref{decovcoupling} to get
\begin{align}
\frac{\mathrm{d} E(r)}{\mathrm{d}r} - 
\Bigg[
(\alpha +1)
\frac{ e'(r)}{e(r)}-\frac{\alpha }{r}
\Bigg]
E(r) &=0,
\end{align}
which gives a relation between the electric field $E(r)$ and the electromagnetic coupling $e(r)^{\alpha +1}$. Then, we have
\begin{align}
E(r) = \ &A \Bigg[\frac{e(r)^{1+\alpha}}{r^{\alpha}}\Bigg].
\end{align}
Here, $\epsilon$ is an integration constant
which controls the strength of the scale dependence,
and which
is thus the called ``running parameter''.
As (\ref{Gr}) shows,
the NEC is a useful tool in order to decrease the number of degrees of freedom of the problem.
\begin{figure*}[ht]
\centering
\includegraphics[width=0.32\textwidth]{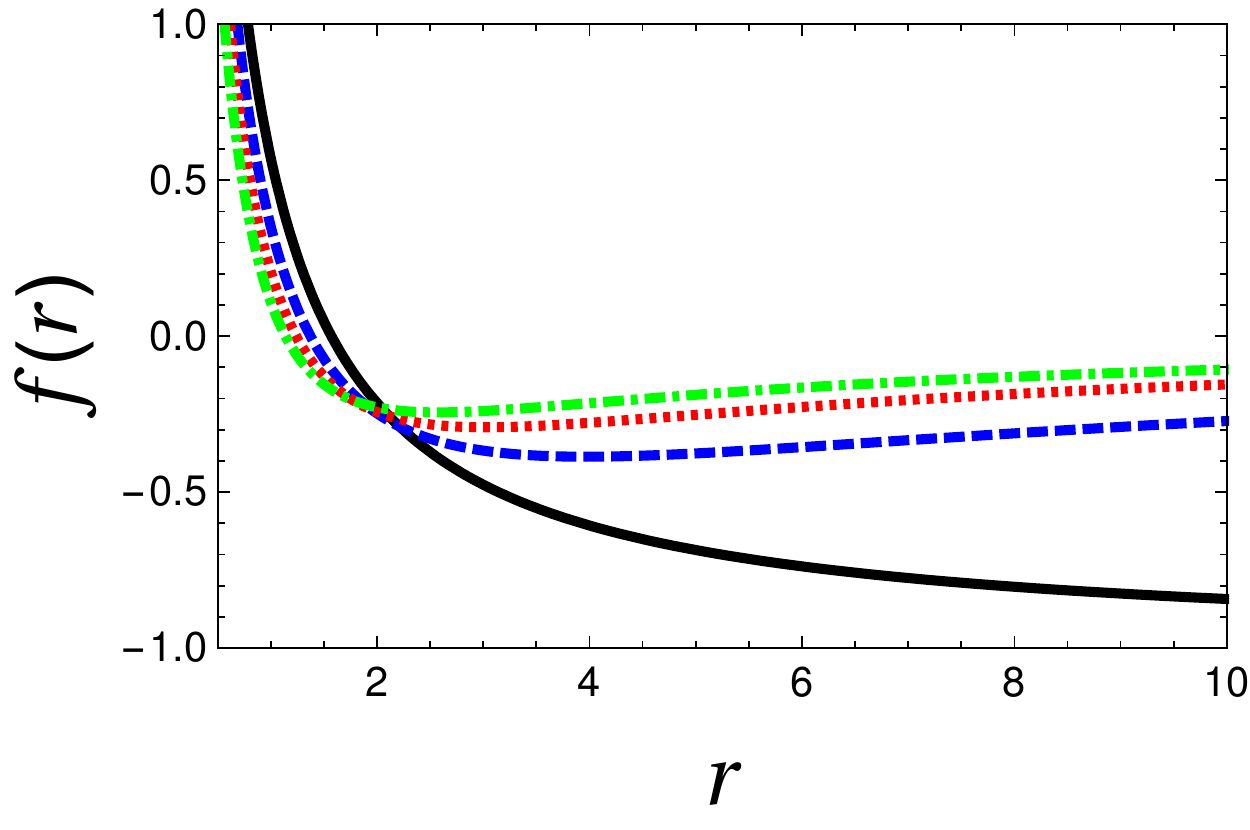}   \
\includegraphics[width=0.32\textwidth]{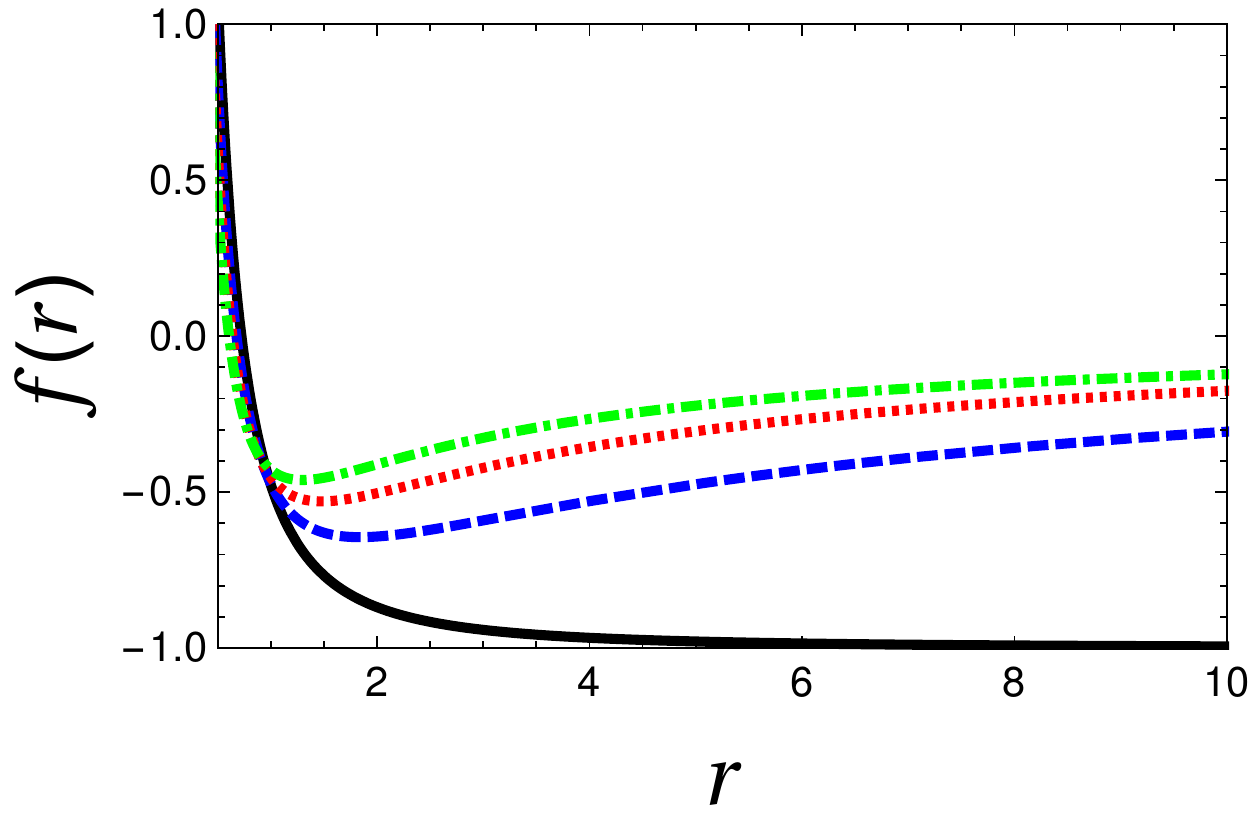}  \
\includegraphics[width=0.32\textwidth]{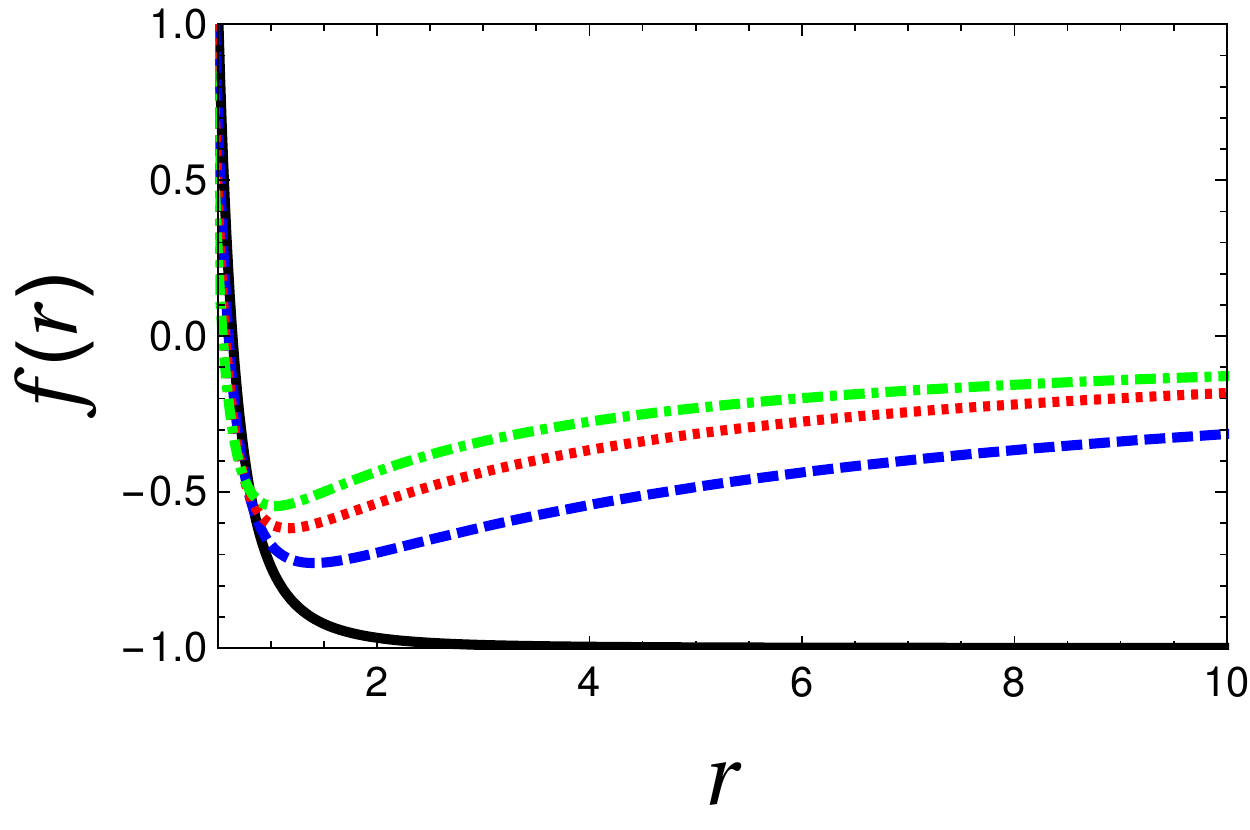} 
\\
\includegraphics[width=0.32\textwidth]{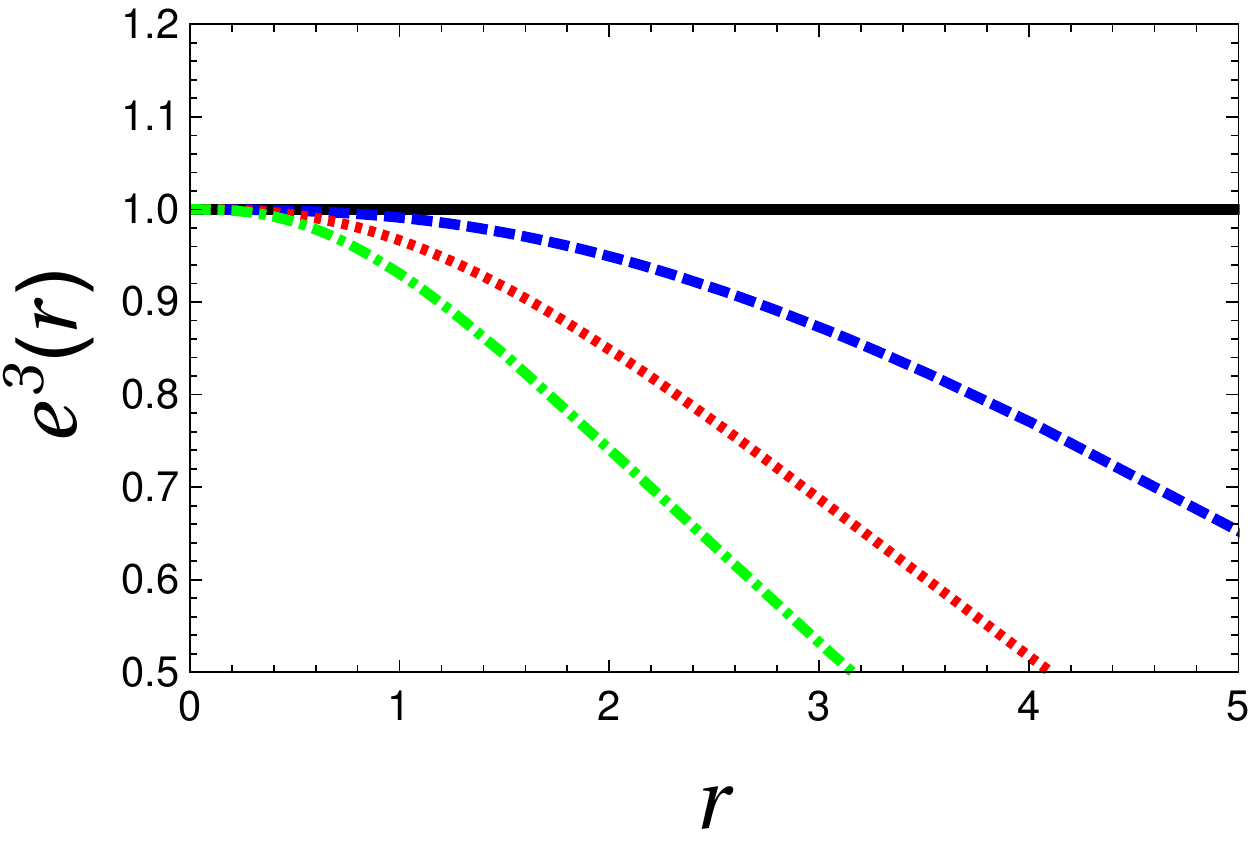}   \
\includegraphics[width=0.32\textwidth]{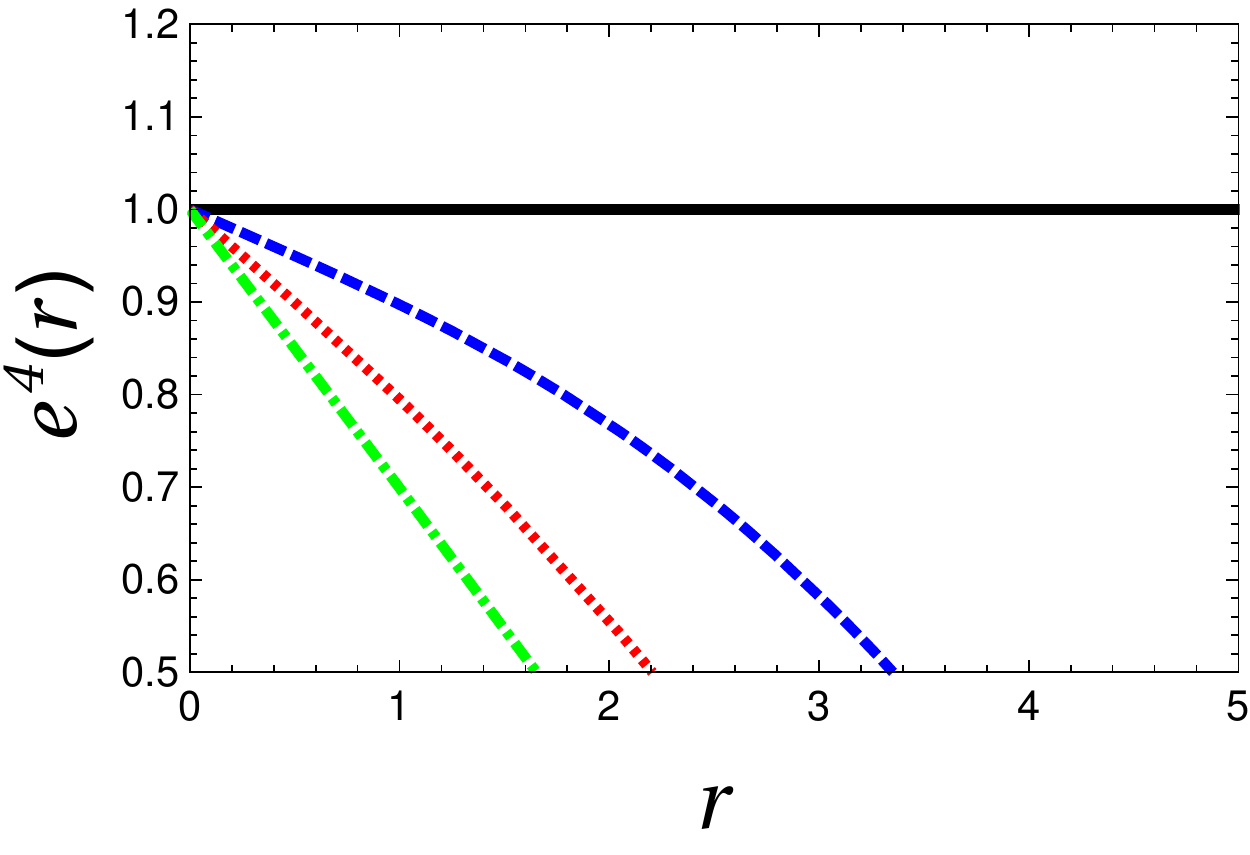}  \
\includegraphics[width=0.32\textwidth]{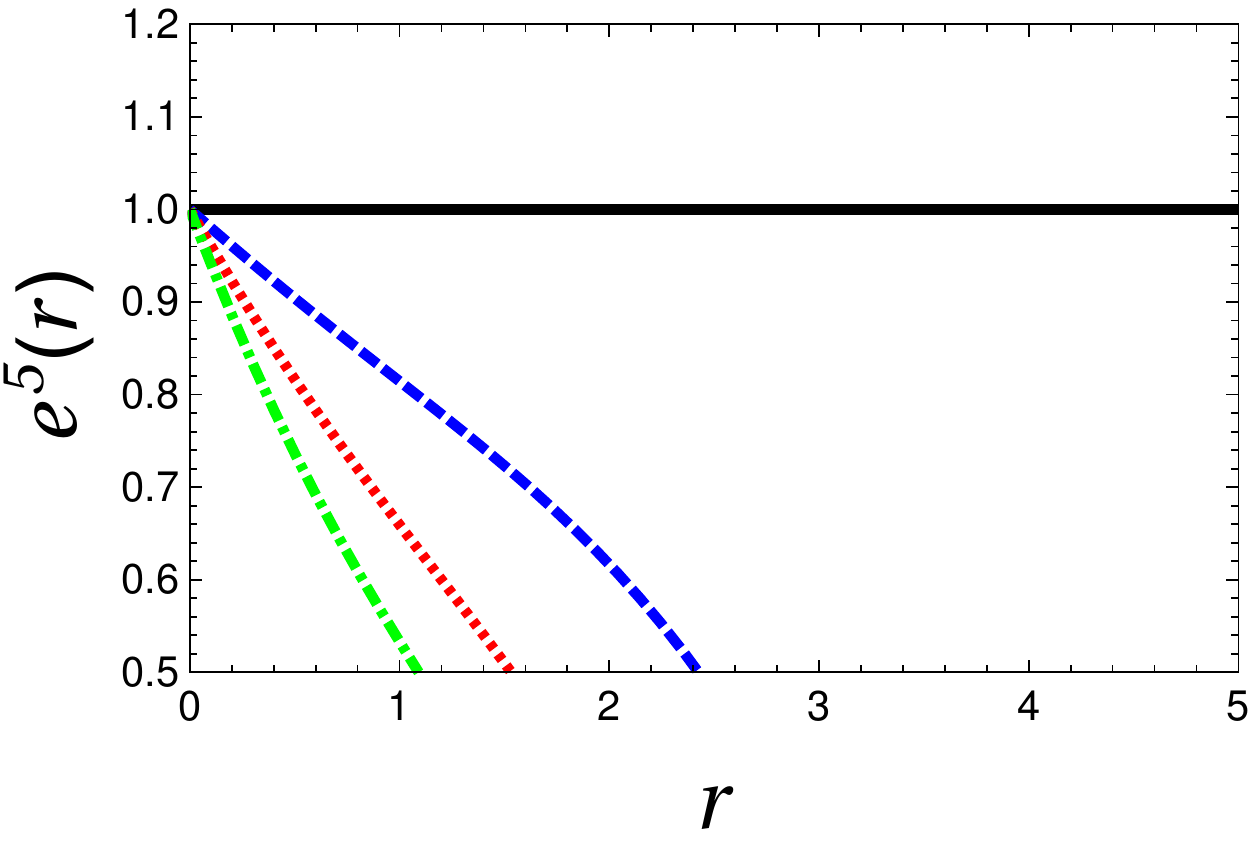} \
\caption{
The lapse function $f(r)$ and the electromagnetic coupling $e(r)^{\alpha+1}$ versus radial coordinate $r$ for  three cases. The first 
line correspond to the lapse function while the second line correspond to the electromagnetic coupling. The first (left), 
second (center) 
and third (right) column 
correspond to the cases $\alpha= \{2,3,4\}$ respectively. We show the classical model (solid black line) and three different cases for each 
figure: i) $\epsilon = 0.1$ (dashed blue line), ii) $\epsilon = 0.2$  (dotted red line) and iii) for $\epsilon = 0.3$  (dotted dashed green line). We have used the set $\{Q_0, M_0, \gamma, G_0\} = \{1,1,1,1/8\}$ in both set of figures. Besides, to complete the scale setting we have used certain $A$ values such as $e_0^{\alpha + 1}$ remains as unity. They are $\{A(\alpha=2),A(\alpha=3),A(\alpha=4)\}= \{0.891, 0.841, 0.812\}$.
}
\label{fig:1}
\end{figure*}
The Einstein field equations give: 
\begin{align}
\begin{split}
 \pi  2^{\frac{7}{2}-\frac{1}{2 \alpha }} 
 \gamma  G_0 A^{\frac{1}{\alpha }+1} & r^{-\alpha } e(r)^{\alpha +1}  \ \ +
\\
& \alpha  (2 r \epsilon +1) f'(r)+2 \alpha  \epsilon  f(r) = 0
\end{split}
\\
\begin{split}
\pi  2^{\frac{7}{2}-\frac{1}{2 \alpha }} \gamma  G_0 A^{\frac{1}{\alpha }+1} & r^{-\alpha } e(r)^{\alpha +1} \ \ - 
\\
&
r \Bigl((r \epsilon +1) f''(r)+2 \epsilon  f'(r)\Bigl) = 0
\end{split}
\end{align}
where the lapse function $f(r)$ and the electromagnetic coupling $e(r)^{\alpha + 1}$
gives the solution: 
\begin{align}
\begin{split}
f(r)  = \ & r^{-\alpha } (1 + \epsilon r)^{-\alpha -1} 
\bigg[
B r + C \Gamma (\alpha -1) r^{\alpha }
\\
& \times \ 
_2\tilde{F}_1(\alpha -1,-\alpha ;\alpha ;-r \epsilon )
\bigg],
\end{split}
\\
\begin{split}
e(r)^{\alpha +1}  = \ & \frac{D}{r (1 + r \epsilon)^{\alpha+2} }
\Bigg[
(1 - \alpha) C r^{\alpha } (1 + 2 \epsilon r)
\\
 &  (1 + \epsilon r)^{\alpha +1}
+ 
\Bigl[
\alpha  (1 + 2 \epsilon r)^2 - 
\\
& 
2 \epsilon r  (2 + \epsilon r) - 1
\Bigl] \times \Bigl[
(\alpha -1) B r + C \ \times
\\
& r^{\alpha }\, _2F_1(\alpha -1,-\alpha ;\alpha ;-r \epsilon )
\Bigl]
\Bigg],
\end{split}
\end{align}
where $D$ is an auxiliary parameter given by
\begin{align}
%\alpha = \ &\frac{1}{1-2\beta},
%\\
D = \ & \frac{2^{\frac{1}{2} \left(\frac{1}{\alpha }-7\right)} \alpha  A^{-\frac{\alpha +1}{\alpha }}}{\pi  (\alpha -1) \gamma  G_0},
\end{align}
and $_2\tilde{F}_1(\ \cdot \ , \ \cdot \ ; \ \cdot \ ; \ \cdot \ )$ is the so-called Hypergeometric Regularized function defined as
follows:
\begin{align}
_2\tilde{F}_1(a,b;c;z) = \ & \sum _{s=0}^{\infty } \frac{1}{\Gamma(c+s)} (a)_s (b)_s \frac{z^s}{s!},
\end{align}
where $(c)_n$ is the (rising) Pochhammer symbol, i.e.
\begin{equation}
(c)_{s} =
\left\{
\begin{array}{ccl}
1
&
\hspace{0.5cm}
\text{if}
\hspace{0.5cm}
&
s=0 
\\
&
&
\\
c(c+1)\cdots (c+s-1)
&
\hspace{0.5cm}
\text{if}
\hspace{0.5cm}
&
s>0
\end{array}
\right.
\end{equation}
Please, note that $_2\tilde{F}_1(a,b;c;z)$ is finite for all finite values of $a$, $b$, $c$, and $z$ as
long as $|z| <1$. Outside the circle $|z| < 1$, the function is defined as the analytic continuation with respect to $z$ of this sum, the parameters $a$, $b$, $c$ held fixed \cite{wolfram}. Besides, the special case $_2\tilde{F}_1(a,b;c;z) = 0$ is forbidden because we assume a non--null $_2\tilde{F}_1$ in the computation of thermodynamic quantities.
In general, the constants are chosen such that the solution matches the classical case when the running
parameter is switched off $\epsilon \rightarrow 0$. However, as the final result depends on the value of the free index $\beta$ (or $\alpha$), we first 
need to take some particular values of these parameters. 
We must emphasize the number of integration constants involved into the problem. Firstly, the scale--dependent gravitational coupling introduce two of them, i.e. $G_0$ and $\epsilon$. This is because we are in the presence of a second order differential equation . The electromagnetic field gives and additional integration constant $A$ whereas the solution for the lapse function implies two additional integration constants $B$ and $C$ (for the same reason as is the gravitational coupling case).
Thus, the integration constant $C$ can be associated with the classical mass of the black hole $M_0$, an the constant $B$ encodes the classical charge $Q_0$. Following Ref. \cite{Gurtug:2010dr} we can set the relation between our integration constants and the classical counterpart as:
\begin{align}
C & \rightarrow \eta M_0 = -8 G_0 M_0(\alpha -1), 
\\
B & \rightarrow  \xi Q_0^{\frac{1+\alpha}{\alpha}} = 
\frac{8 \pi G_0 }{(\alpha -1) \alpha }
 Q_0^{\frac{1+\alpha}{\alpha}}.
\end{align}
Thus, we have a link between the usual solution and the scale-dependent one.
We emphasize that $M_0$ is the classical mass, not to be confused with the mass of the scale-dependent black hole. The $M_0$ identification is made when we take the limit $ \epsilon \rightarrow 0 $, since the scale-dependent solution tends to the classical one in that limit.
According to the previous expressions we observe that the parameters of the theory depend on the the power $\beta$ of the theory, in total agreement with the classical one.
Furthermore, an important check is that our solution reproduces the results of the classical theory in the limit $\epsilon \rightarrow 0$, i.e.
\begin{align}
\begin{split}
\lim_{\epsilon \rightarrow 0} G(r) &= G_0, 
\\
\lim_{\epsilon \rightarrow 0} E(r) &= E_0(r) = A \Bigg[\frac{B D (1 - \alpha)^2}{r^{\alpha }}\Bigg] ,
 \\
\lim_{\epsilon \rightarrow 0} f(r) &= f_0(r) = - \frac{C}{1 - \alpha} + B r^{1-\alpha },
\\
\lim_{\epsilon \rightarrow 0} e(r)^{\alpha +1} &= e_ 0^{\alpha+1} = B D (1 - \alpha)^2 .
\end{split}
\end{align}
where the parameters $\{A, B, C, D\}$ have fixed values according to \cite{Gurtug:2010dr}
in terms of their meaning in the absence of scale dependence \cite{angel}.
The scale dependent scenario introduces small corrections 
to the fixed-scale background, as can be easily seen by
\begin{align}
G(r) &\approx G_0 \Bigl[1 - \epsilon r \Bigl]
+
\ \mathcal{O}(\epsilon^2),
\\
E(r) &\approx E_0(r)
\Bigl[
1-(\alpha -2) r \epsilon
\Bigl]
+
\ \mathcal{O}(\epsilon^2),
\\
f(r) &\approx f_0(r) +  
\left[
\frac{2 C r}{1-\alpha}
-
\frac{(\alpha +1) B}{r^{\alpha-2 }} 
\right]
\epsilon
+
 \mathcal{O}(\epsilon^2),
\\
e(r)^{\alpha + 1} &\approx e_0^{\alpha + 1} 
\Bigl[
1-(\alpha -2) r \epsilon
\Bigl]
+
\ \mathcal{O}(\epsilon^2).
\end{align}
Finally, we should remark that certain values of the power $\alpha$ are forbidden. As our solution must be valid indeed in the classical case, the 
first step is to analyze this solution.  In order to avoid singularities in the classical lapse function, $\alpha = 1$ is excluded.  Following the same line of thought, the 
scale--dependent lapse function
implies that $\alpha \neq 0$ is forbidden. Hence,  the two parameters $\alpha = \{0 , 1\}$ are not permitted. Besides, all complex numbers 
except the non-positive integers (where the function has simple poles), are, in principle, possible. Despite that, we will focus on cases where $\alpha \geq 2$. In Fig \ref{fig:1} we observe the behaviour of the lapse function and the electromagnetic coupling for different values of the parameter $\beta$.

%%%%%%%%%%%%%%%%%%
\subsection{Asymptotic behaviour}
The asymptotic behaviour will be studied using two curvature invariants, i.e. the Ricci scalar as well as the Kretschmann scalar. 
These invariants give information related to possible divergences, which is crucial for the diagnostic of our solution. To complete the analysis, we will include in our discussion the coordinate dependent  (not invariant) asymptotic lapse function.
Given the metric function \eqref{metric}, the 
%Ricci scalar is given by
scalars are given by
\begin{align}
R & = -f''(r) - 2 \frac{f'(r)}{r},
\end{align}
\begin{align}
\mathcal{K} \equiv \ & R_{\mu \nu \alpha \beta}R^{\mu \nu \alpha \beta} = f''(r)^2 + 2 \Bigg(\frac{f'(r)}{r}\Bigg)^2,
\end{align}
which, in our particular case, take the form:
\begin{align}
\begin{split}
R = &\frac{C (\alpha +2 \alpha  r \epsilon -2)}{r^2 (r \epsilon +1)^2} - r^{-\alpha -2} (r \epsilon +1)^{-\alpha -3} \ \times 
\\
&\Bigg[
(\alpha +2 \alpha  r \epsilon )^2-\alpha  (2 r \epsilon  (r \epsilon +4)+3)+2
\Bigg] \ \times
\\
&B r+C \Gamma (\alpha -1) r^{\alpha } \, _2\tilde{F}_1(\alpha -1,-\alpha ;\alpha ;-r \epsilon ),
\end{split}
\end{align}
%and
\begin{align}
\begin{split}
\mathcal{K} &=   \frac{r^{-2 (\alpha +2)} (r \epsilon +1)^{-2 (\alpha +3)}}{\Gamma (\alpha )^2}
\Bigg[
2 (r \epsilon +1)^2 
\Bigg{ \{ }
B r 
\\
& 
\times \Gamma (\alpha )
(\alpha +2 \alpha  r \epsilon -1) -C \Gamma (\alpha -1) r^{\alpha }
\Bigg(
(\alpha -1) 
\\
& \times  (r \epsilon +1)^{\alpha +1} -\Gamma (\alpha ) (\alpha +2 \alpha  r \epsilon -1)
\\
& \times \, _2\tilde{F}_1(\alpha -1,-\alpha ;\alpha ;-r \epsilon )
\Bigg)
\Bigg{ \} }^2
+
\Bigg{ \{ }
B r \Gamma (\alpha )
\\
&
\times
\Bigg[
\alpha \left(-2 r^2 \epsilon ^2+2 r \epsilon +1\right)-(\alpha +2 \alpha  r \epsilon )^2+2 r \epsilon
\Bigg]
\\
& + 
C \Gamma (\alpha -1) r^{\alpha } 
\Bigg[
\Gamma (\alpha ) 
(\alpha  
\left(-2 r^2 \epsilon ^2+2 r \epsilon +1
\right) 
\\
&
-  (\alpha +2 \alpha  r \epsilon )^2 + 2 r \epsilon 
) 
\, _2\tilde{F}_1(\alpha -1,-\alpha ;\alpha ;-r \epsilon )
\\
&
+ 
(\alpha -1) (\alpha +2 \alpha  r \epsilon +2 r \epsilon ) (r \epsilon +1)^{\alpha +1}
\Bigg]
\Bigg{ \} }^2
\Bigg].
\end{split}
\end{align}
Thus, the classical values for the scalars are 
%again recovered requiring $\epsilon \rightarrow 0$, which reads
\begin{align}
R_0 \equiv \ & \lim_{\epsilon \rightarrow 0} R = \frac{(\alpha -2) (1- \alpha) B }{r^{\alpha + 1}},
\end{align}
\begin{align}
\mathcal{K}_0 \equiv \lim_{\epsilon \rightarrow 0}\mathcal{K} = \frac{(\alpha -1)^2 \left(\alpha ^2+2\right) B^2}{r^{2 (\alpha +1)}} .
\end{align}

%%%%%%%%%%%%%%%%%%%%%%%%%
\subsubsection{Asymptotics for $r \rightarrow 0$}
%%%%%%%%%%%%%%%%%%%%%%%%%
First, the lapse function in this regime is given by
\begin{align}
f(r \rightarrow 0) = f_0(r) -\frac{2 C \epsilon r }{\alpha -1}
+ \mathcal{O}(r^2),
\end{align}
whereas the invariants take the form:
\begin{align}
R(r \rightarrow 0) = \ & R_0 + 
\frac{4 C \epsilon }{(\alpha -1) r} 
+ 
\mathcal{O}(r^{-\alpha }),
\end{align}
\begin{align}
\mathcal{K}(r \rightarrow 0) = \ & 
\mathcal{K}_0 + 8 B C \epsilon r^{-(\alpha + 2) }
 +
\mathcal{O}\Bigl(r^{-(1 + 2 \alpha)}\Bigl).
\end{align}
\noindent
We see that the scalars have singularities in the scale dependent scenario, i.e.  when we include the running of the coupling constants, just like their scale-independent counterpart.
It would be interesting to investigate how and to which extent those singularities could be cancelled by an additional contribution from the effective stress energy tensor as discussed in \cite{Contreras:2017eza}.

%%%%%%%%%%%%%%%%%%%%%%%%%
\subsubsection{Asymptotics for $r \rightarrow \infty$}
%%%%%%%%%%%%%%%%%%%%%%%%%
As before, it is very useful compute the lapse function for this regime, i.e. 
\begin{align}
\begin{split}
f(r \rightarrow \infty) = \ &
\frac{r^{-\alpha }}{ (r \epsilon +1)^{\alpha + 1}}
\Bigg[
Br 
+
\frac{1}{2} C \epsilon ^{\alpha } r^{2 \alpha }  \ \times
\\
&
\Bigg[
\frac{2}{2 \alpha -1} 
\ +
\frac{\alpha }{(\alpha -1) r \epsilon }
\Bigg] + C r \epsilon ^{1-\alpha } \ \times
\\
&
\frac{ \Gamma (1-2 \alpha ) \Gamma (\alpha -1)}{\Gamma (-\alpha )}
\Bigg] \ + \ \mathcal{O}\bigg(\frac{1}{r}\bigg)^2
\end{split}
\end{align}
Besides, we have that the Ricci scalar can be written  up to zeroth order as:
\begin{align}
\begin{split}
R(r \rightarrow \infty) = 
&
\frac{1}{(r \epsilon +1)^{\alpha +3}}
\Bigg[
R_0(r \rightarrow \infty) - 4 (\alpha -2) 
\\
&\times \alpha  B \epsilon  r^{-\alpha }
+
2 (1-2 \alpha ) \alpha  B \epsilon ^2 r^{1-\alpha } \ -
\\
& \alpha  C \epsilon ^{\alpha +1} r^{\alpha -1} 
\left[\frac{4 \alpha ^3 - 11 \alpha + 8}{2 \alpha ^2-3 \alpha +1} + 
2 r \epsilon \right]
\\
&\frac{-C \epsilon ^{1-\alpha }}{r^{1+\alpha} \Gamma (-\alpha )}
\Bigl(
(\alpha +2 \alpha  r \epsilon )^2-\alpha  (2 r \epsilon  
(r \epsilon 
\\
& + 4)
+3)+2
\Bigl)
 \Gamma (1-2 \alpha ) \Gamma (\alpha -1)
\Bigg]
\end{split}
\end{align}
The Kretschmann scalar has a complicated expansion,
and it is avoided for simplicity. It is remarkable that the invariants in that limit maintain the singularity present in the classical theory.
\begin{figure*}[ht]
\centering
\includegraphics[width=0.32\textwidth]{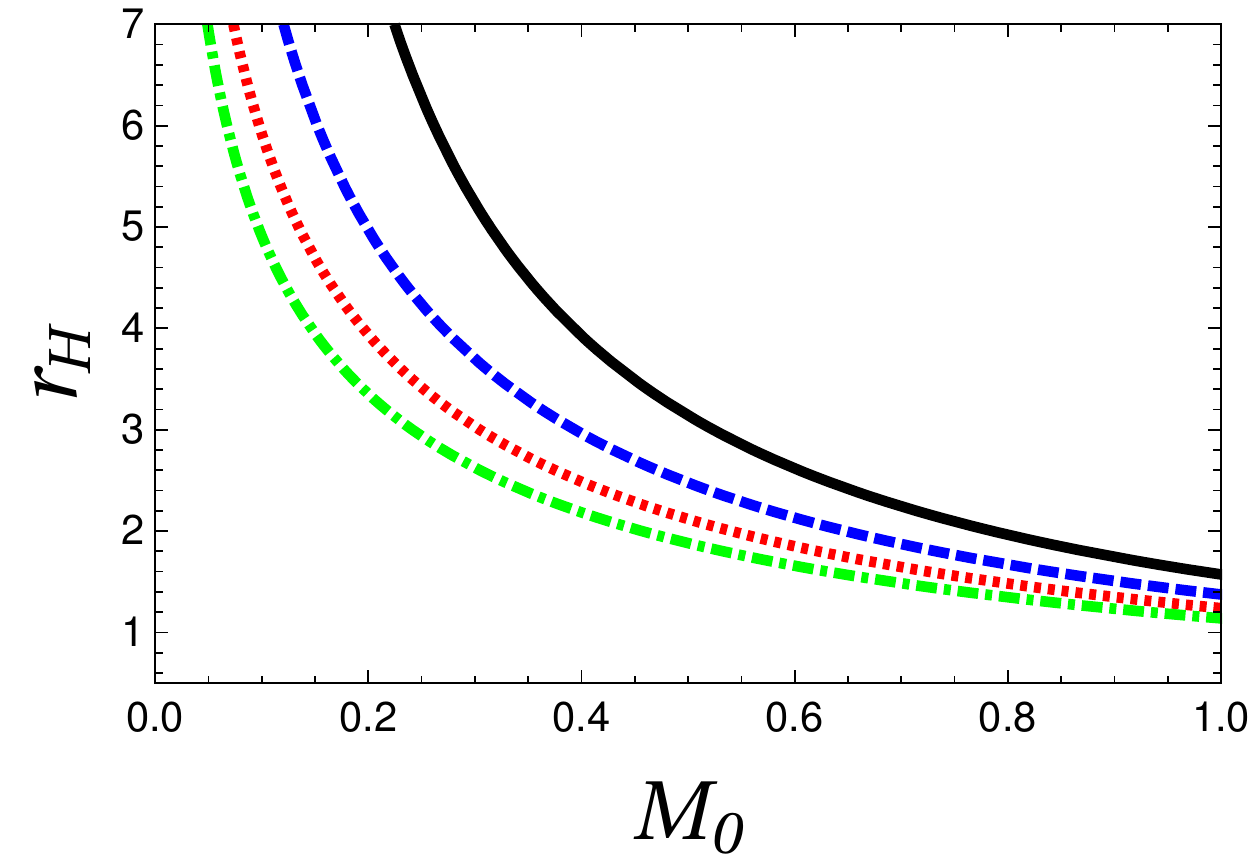}   \
\includegraphics[width=0.32\textwidth]{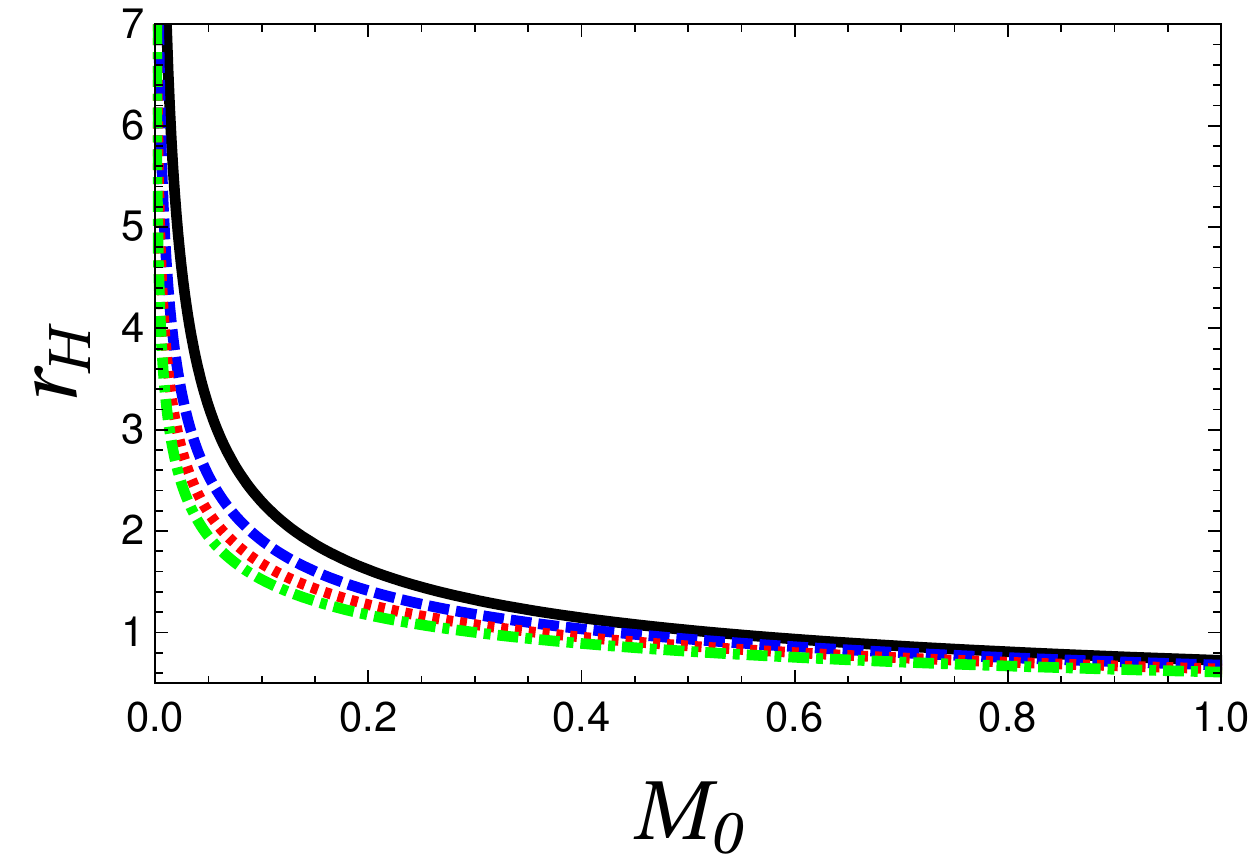}  \
\includegraphics[width=0.32\textwidth]{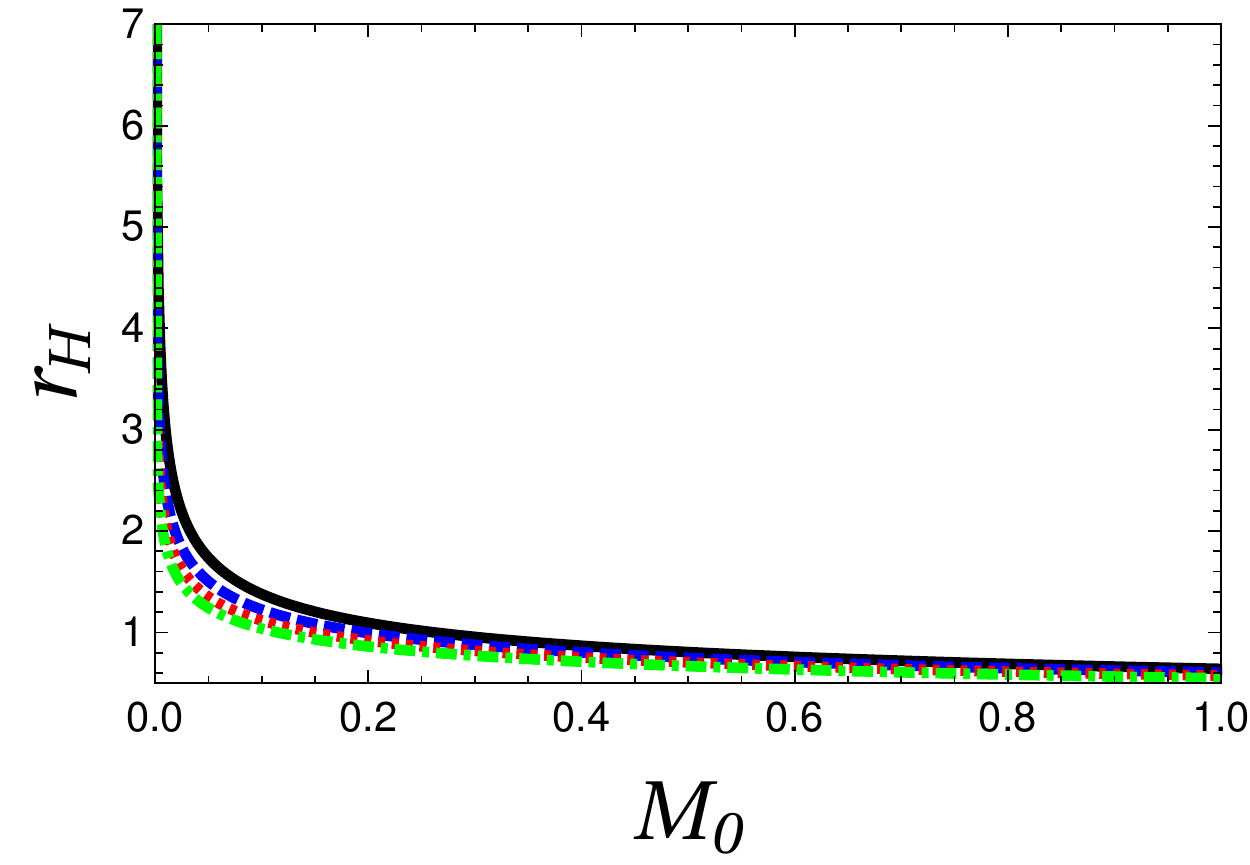} 
\caption{
The evolution of event horizon $r_H$ versus the classical black hole mass $M_0$ for three cases. The first (left), second (center) and third (right) column correspond to the cases $\alpha= \{2,3,4\}$, respectively. We show the classical model (solid black line) and three different cases for each figure: i) $\epsilon = 0.1$ (dashed blue line), ii) $\epsilon = 0.2$  (dotted red line) and iii) for $\epsilon = 0.3$  (dotted dashed green line). We have used the set $\{Q_0, G_0\} = \{1,1/8\}$ in all set of figures.
}
\label{fig:2}
\end{figure*}

\subsection{Horizons}
The event horizon is given when the lapse function vanishes, i.e. $f(r_H)=0$.
Given the functional structure of $f(r)$, it is required to select a certain value of the index $\beta$ (or $\alpha$). 
Note that the effect of scale dependence ($\epsilon \neq 0$) can be understood as a non-trivial deviation from the classical solution ($\epsilon =0$).
As we commented before, we will focus on models where $\alpha \geq 2$. The corresponding lapse functions in that regime has a polynomial structure and the roots usually have a complex form. As a benchmark point, we will revisit the solution for $\alpha =2$ which was previously 
discussed in Ref. \cite{previous}. Note that, although we are able to produce physical solutions for $\alpha \geq 2$, only a single case will be shown here explicitly. Fig. \ref{fig:2} show the behavior of that solution plus two additional 
cases assuming $\alpha = \{3,4\}$. 
The scale dependent lapse function $f(r;\alpha)$ is, for $\alpha =2$,
\begin{align}
&f\left(r;2 \right) = \  \frac{3 B+C r (r \epsilon  (r \epsilon +3)+3)}{3 r (r \epsilon +1)^3},
\end{align} 
whereas the corresponding classical solution is:
\begin{align}
f_0(r) & = C +\frac{B}{r}.
\end{align}
In order to connect the classical with the scale-dependent counterpart, we compute the classical horizon, i.e. $r_0 = -B/C$.
Then we obtain the scale--dependent horizon $r_{H}(\epsilon;\alpha)$ using the classical value, as
\begin{align}
&r_H\left(\epsilon;2\right) = \ - \frac{1}{\epsilon} \Bigg[
1 -(1 + 3 \epsilon r_0)^{1/3} 
\Bigg].
\end{align}
Finally, we recover the classical case expanding the solution for small values of $\epsilon$, that is
\begin{align}\label{eqrHexp}
r_H(\epsilon;2) \approx r_0 \Bigl[1- \epsilon r_0 + {\mathcal{O}}(\epsilon^2) \Bigl].
\end{align}
One notes that the horizon radius in the scale dependent scenario $r_H$ is 
reduced with respect to its classical counterpart $r_0$, this effect can also be appreciated from
the graphical analysis in figure \ref{fig:2}.

\subsection{Thermodynamic properties}

\begin{figure*}[ht]
\centering
\includegraphics[width=0.32\textwidth]{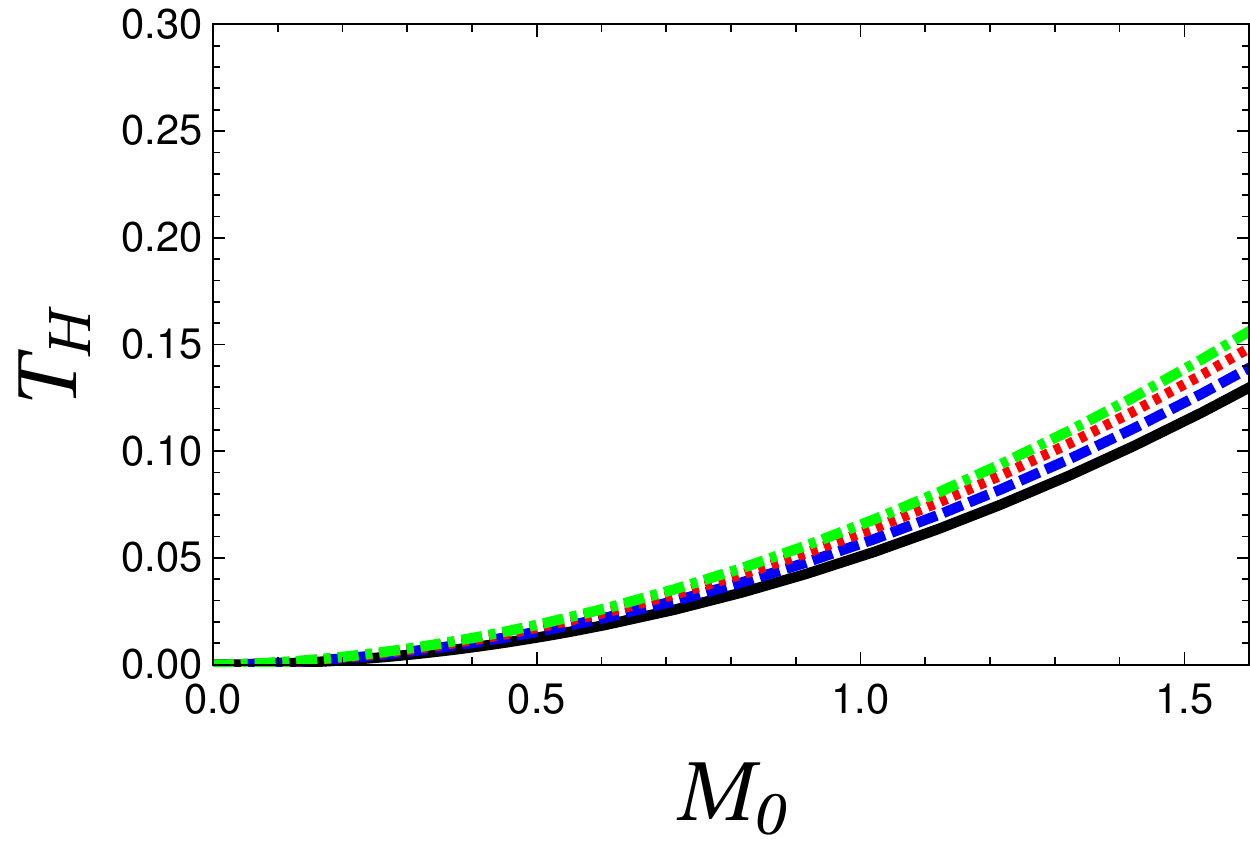}  \
\includegraphics[width=0.32\textwidth]{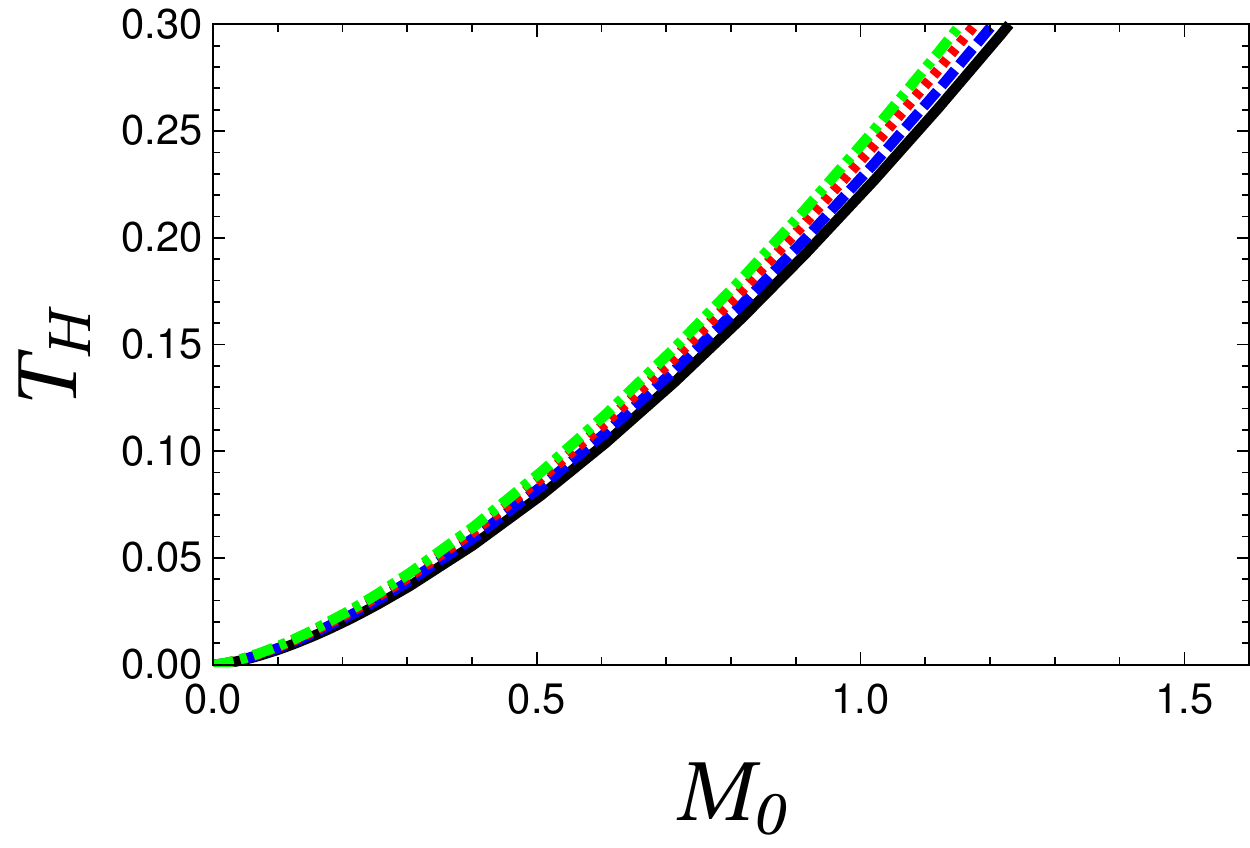}  \
\includegraphics[width=0.32\textwidth]{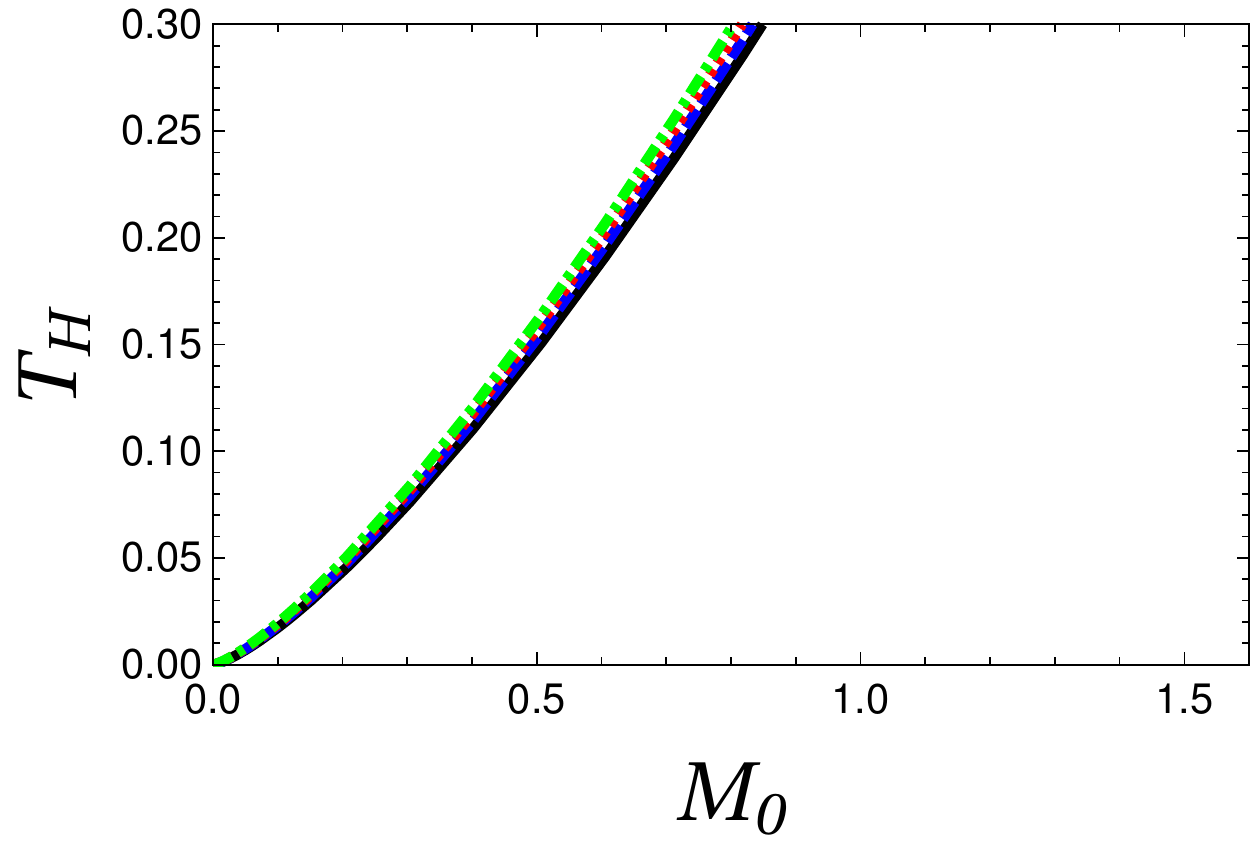} 
\\
\includegraphics[width=0.32\textwidth]{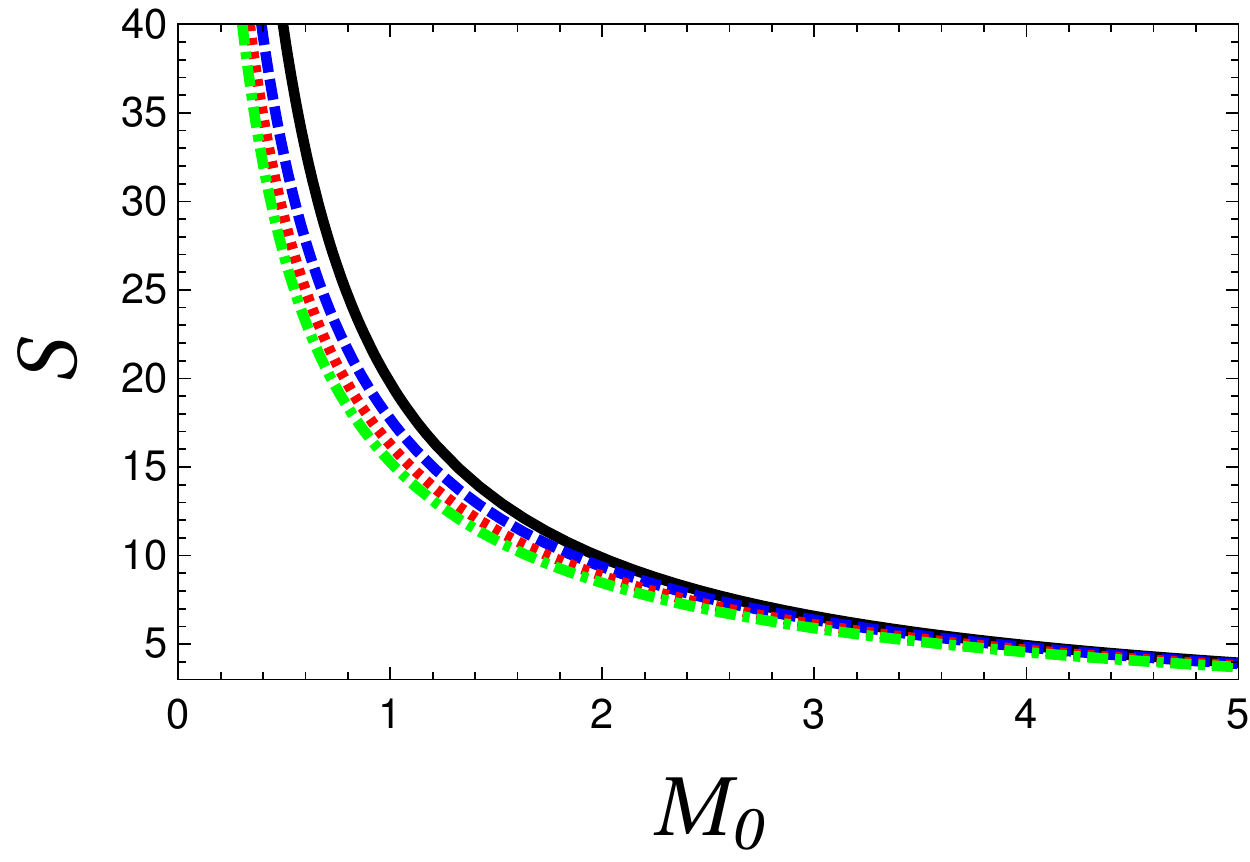}  \
\includegraphics[width=0.32\textwidth]{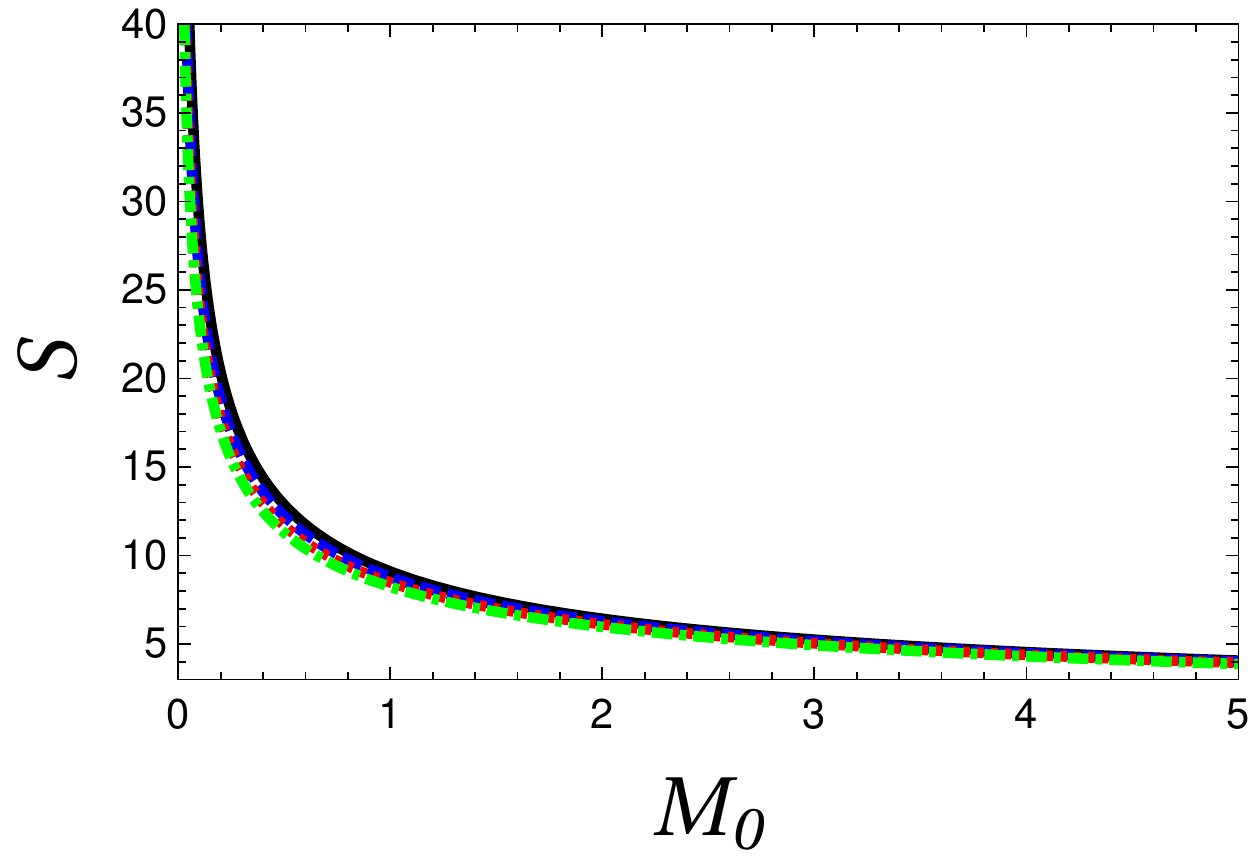}  \
\includegraphics[width=0.32\textwidth]{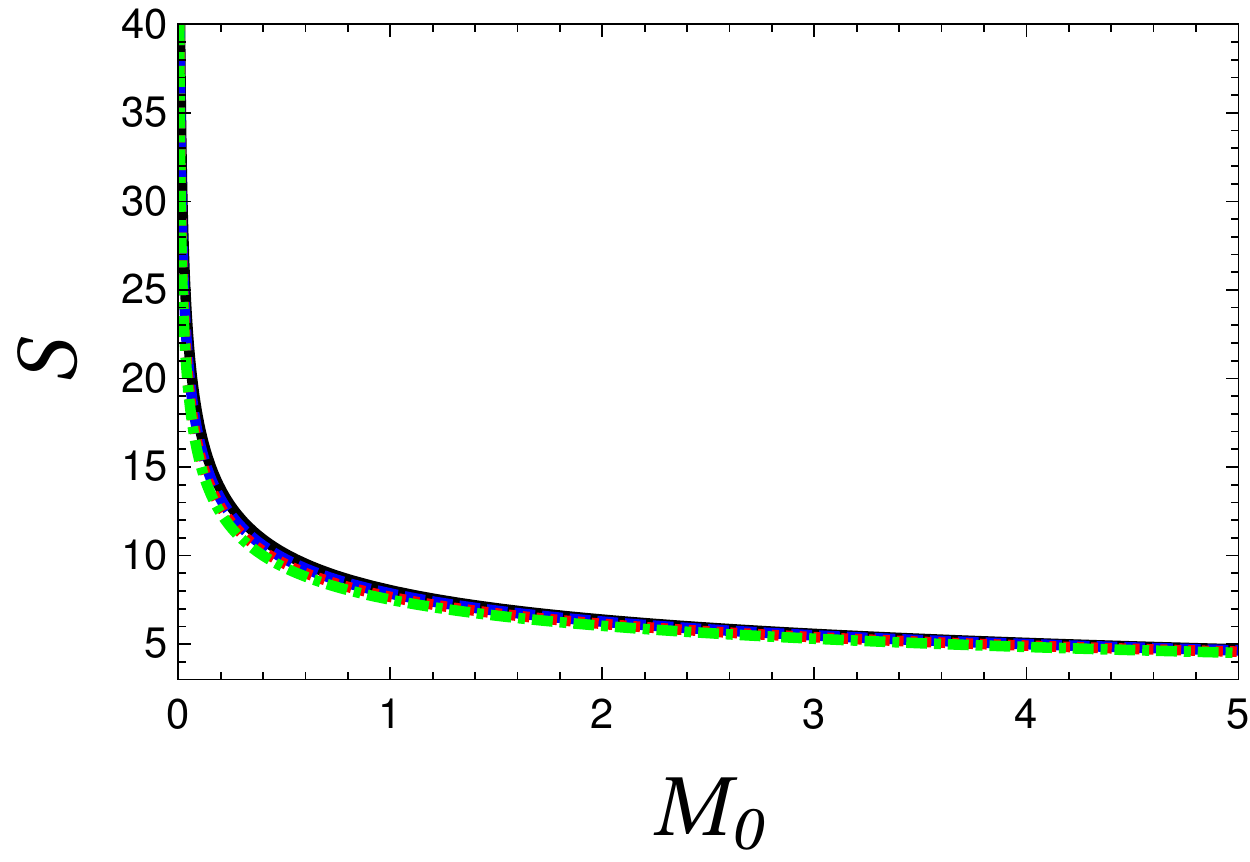}  \
\caption{
The Hawking temperature $T_H$ and Bekenstein Hawking entropy $S$ versus the classical mass $M_0$ for  three cases. The first
line correspond to the Hawking temperature while the second line correspond to the Bekenstein Hawking entropy. The first (left), second (center) and 
third (right) column correspond to the cases $\alpha= \{2,3,4\}$, respectively. We show the classical model (solid black line) and three different cases 
for each figure:  i) $\epsilon = 1$ (dashed blue line), ii) $\epsilon = 2$  (dotted red line) and iii) for $\epsilon = 3$  (dotted dashed green line). We have used the set $G_0 = 1/8$ in all set of figures.
}
\label{fig:3}
\end{figure*}

The horizon structure provides  the required information in order to obtain thermodynamic properties like temperature and entropy.
On one hand, the Hawking temperature for the ansatz (\ref{metric}) is given by
\begin{align}
T_H(r_H) &= \frac{1}{4 \pi} \Bigg|\lim_{r\rightarrow r_H} \frac{\partial_r g_{tt}}{\sqrt{-g_{tt}g_{rr}}} \Bigg|,
\end{align}
i.e. 
\begin{align}
T_H(r_H) &= \frac{1}{4 \pi}\Bigg|\frac{C}{r_H(1 + \epsilon r_H)}\Bigg|.
\end{align}
One notes that the functional structure of Hawking temperature remains invariant under changes of the parameter $\alpha$.  In addition, note that we recover the classical solution after
demanding $\epsilon \rightarrow 0$.
Taking into account the scale-dependent philosophy, the solution can be expanded around $\epsilon = 0$ 
\begin{align}
T_H(r_H) &\approx T_0(r_0) \Bigl|1 + \epsilon r_0  + \mathcal{O}(\epsilon^2)\Bigl|,
\end{align}
where $r_0$ is the classical horizon. 
Clearly, the classical result $T_0$ is recovered for $\epsilon \rightarrow 0$.
In Figure \ref{fig:3} we show the scale--dependent temperature which takes into account the running coupling effect.
\noindent 
On the other hand, the Bekenstein-Hawking entropy for 
Brans-Dicke type theories is known to be
\begin{align}
S &= \frac{1}{4} \oint  {\mathrm {d}}x \frac{\sqrt{h}}{G(x)},
\end{align}
where $h_{ij}$ is the induced metric at the horizon. In presence of circularly symmetric solution and taking advantage of the fact that $G(x) = G(r_H )$ is 
constant along the
horizon, this integral takes the form \cite{angel,Rincon:2017ypd}
\begin{align}\label{eqSs}
S &=\frac{\mathcal{A}_H(r_H)}{4G(r_H)} = S_0(r_H)(1 + \epsilon r_H).
\end{align}
Note that the relation (\ref{eqSs}) naively  suggests that the entropy  increases for increasing $\epsilon$,
this effect is however overcompensated by the decrease in the black hole horizon $r_H$ as it can be appreciated from e.g. (\ref{eqrHexp}).
In the lower part of Figure \ref{fig:3}  we show the entropy for the generalized (2+1)-dimensional Einstein-power-Maxwell scale dependent black hole. It is evident that the running effect is important when $\epsilon r$ is large, however, we remain small values of the parameter $\epsilon$ following the idea that quantum correction should be just small corrections to the classical solution.
\noindent To conclude, the heat capacity can be obtained from the usual relation
\begin{align}
C_Q &= T \ \frac{\partial S}{\partial T}\Bigg|_{Q} ,
\end{align}
which gives
\begin{align}\label{Heat}
C_Q &= -S_0(r_H)(1 + \epsilon r_H ),
\end{align}
where we have used the chain rule through the relation $\partial S/ \partial T = (\partial S/ \partial r_H)(\partial r_H/\partial T)$. It is important to note that solution \eqref{Heat} is an exact result and, indeed, gives us the classical solution after demanding $\epsilon \rightarrow 0$.
Besides, due to a weak $\epsilon$ dependence it was necessary to 
plot all the figures with very large values of $\epsilon$ in order to generate an appreciable effect. The scale 
dependent effect is notoriously small for those quantities.

Regarding the Smarr formula and the first law of black hole mechanics, we remark a couple of facts first: given that we work in the framework of non-linear electrodynamics, we expect to have a modified relation compared to Maxwell's linear theory, as it has been shown in \cite{Balart:2017dzt}. Furthermore, the Smarr formula requires knowledge of the total mass $M$, which unfortunately in the present work is unknown. In spite of that, to get some insight into the underlying physics, we take the case where $\alpha =2$ to exemplify how the new Smarr-like relation looks like. It is straightforward to check that in the classical theory one obtains:
\begin{align}
M_0 = T_0 S_0,
\end{align} 
while in the scale-dependent scenario, to leading order in $\epsilon$, we find
\begin{align}
M \approx M_0 \approx T_H S_H - \epsilon \left(\frac{1}{2}\pi Q_0^{3/2}\right).
\end{align}
Note that in the weak regime $(r \epsilon \ll 1)$ we have approximated the total mass as the classical one. This should be a good approximation, as we expect that any deviations from the classical value will be small. A more detailed analysis of the Smarr formula is beyond the purpose of this paper, and we hope to be able to address this issue in more detail in a future work.

Before we conclude our work a final comment is in order here. The no-go theorem of \cite{Ida:2000jh}, which links the existence of smooth black hole horizons to the presence of a negative cosmological constant, does not apply in the given case. First, the theorem is based on unmodified classical Einstein Field equations, which is not the case in scale-dependent scenarios. Second, the no-go theorem assumes the dominant energy condition which is not part of our assumptions. Instead, we take advantage of the so-called null energy condition. Furthermore, and most importantly, given the solutions previously presented one can check that they do have smooth horizons and well behaved asymptotic spacetimes, and therefore they are black holes. Note that even the classical solution in \cite{Gurtug:2010dr} was shown to be a black hole in this sense, even for a vanishing cosmological constant.

\section{Conclusions} \label{Conclusions}

In the present article we have studied the effect of scale dependent couplings on charged black holes in the presence of three-dimensional 
Einstein-power-Maxwell non-linear electrodynamics for any value of the power parameter, extending and generalizing a previous work. 
First we presented the model and the classical black hole solution assuming static circular symmetry, and then we allowed for a scale dependence of 
the couplings, both the electromagnetic and the gravitational one. We solved the corresponding effective field equations applying the same formalism 
already used in our previous work, namely by imposing the "null energy condition". Black hole properties, such as horizon structure, Hawking temperature, 
Bekenstein-Hawking entropy as well as asymptotic properties, are discussed in detail. 
In order to show how the scale--dependent scenario modifies the classical solution, we have considered three different 
 benchmark cases taking $\alpha = \{2, 3, 4 \}$ which are shown in Fig. \ref{fig:1}, \ref{fig:2} and \ref{fig:3}. The aforementioned solutions have a 
 managable mathematical structure which allows to obtain analytical expressions for the physical quantities.
The solutions obtained in this work and our main 
numerical results show that the scale--dependent scenario allows us to induce deviations from classical black hole solutions, confirming a 
result already reported in \cite{Gurtug:2010dr}. In particular, it is worth mentioning that the behavior of the electromagnetic coupling depends drasstically on the choice of the parameter $\alpha$. 
Regarding the basic black hole  properties, we have found that for a fixed
classical black hole mass, the Hawking temperature
increases with $\epsilon$, while both the event horizon radius and the Bekenstein-Hawking entropy decrease when the strength of the scale dependence 
increases. Our findings imply that quantum corrections may have an remarcable effect, i.e. the black hole becomes hotter and at the same time loses less
information compared to its classical counterpart. This is in agreement with the findings in 
\cite{Bonanno:2000ep,Bonanno:2006eu,Reuter:2006rg,Reuter:2010xb,Falls:2012nd
,Cai:2010zh,Becker:2012js,Becker:2012jx,Koch:2013owa,Koch:2013rwa,Ward:2006vw,Burschil:2009va,Falls:2010he,Koch:2014cqa,Bonanno:2016dyv}.
Finally, it is well-known that a black hole, viewed as a thermodynamical system, is locally stable if its heat
capacity is positive \cite{Dehghani:2016agl}. We have found that the black holes studied here are unstable ($C_Q < 0$), both classically and
in the scale dependent scenario. 
To conclude, our results allow us to gain a solid understanding of the most important modifications that a possible scale dependence would imply for the Einstein-Maxwell black holes of arbitrary power in $2+1$ dimensions.
\begin{acknowledgements}
The author A.R. was supported by the CONICYT-PCHA/\- Doctorado Nacional/2015-21151658.
The author P.B. was supported by the Faculty of Science and Vicerrector\'{\i}a de Investigaciones of Universidad de los
Andes, Bogot\'a, Colombia. 
The author B.K. was supported by the Fondecyt 1161150.
The author G.P. thanks the Funda\c c\~ao para a Ci\^encia e Tecnologia (FCT), Portugal, for the financial support to the Center for Astrophysics and Gravitation-CENTRA, Instituto Superior T\'ecnico, Universidade de Lisboa, through the Grant No. UID/FIS/00099/2013.
\end{acknowledgements}

%\bibliography{refsBTZ}   % name your BibTeX data base

\bibliographystyle{unsrt}         % Orden de aparicion: Angel

\end{document}